\documentclass[10pt, twocolumn]{IEEEtran}
\usepackage{fancyhdr}
\usepackage{epsfig}
\usepackage{threeparttable}
\usepackage{epsf,epsfig}
\usepackage{amsthm}
\usepackage{amsmath}
\usepackage{amssymb}
\usepackage{amsfonts}
\usepackage[noadjust]{cite}
\usepackage{dsfont}
\usepackage{subfigure}
\usepackage{color} 
\usepackage{enumerate}
\usepackage{commath}

\def\({\left(}
\def\){\right)}

\def\[{\left[}
\def\]{\right]}

\begin{document}

\title{Resource Allocation with Reverse Pricing for Communication Networks}

\author{\IEEEauthorblockN{Sang Yeob Jung and Seong-Lyun Kim}\\
\IEEEauthorblockA{School of Electrical and Electronic Engineering, Yonsei University\\
50 Yonsei-Ro, Seodaemun-Gu, Seoul 120-749, Korea\\
Email: \{syjung, slkim\}@ramo.yonsei.ac.kr}}

\maketitle

\begin{abstract}

Reverse pricing has been recognized as an effective tool to handle demand uncertainty in the travel industry (e.g., airlines and hotels).
To investigate its viability for communication networks, 
we study the practical limitations of (operator-driven) time-dependent pricing that has been recently introduced, taking into account demand uncertainty.
Compared to (operator-driven) time-dependent pricing, 
we show that the proposed pricing scheme can achieve ``triple-win'' solutions: an increase in the total average revenue of the operator; 
higher average resource utilization efficiency; and an increment in the total average payoff of the users. 
Our findings provide a new outlook on resource allocation, and design guidelines for adopting the reverse pricing scheme.
\end{abstract}

\begin{keywords}
Reverse pricing, communication networks, revenue management, resource allocation, demand uncertainty.
\end{keywords}

\section{Introduction}

As a tool for balancing capacity and demand, 
pricing has played a decisive role in the design and management of modern communication networks  \cite{Kelly98}. 
In order to achieve various network objectives (i.e., ``optimization-oriented'' for maximum system performance or ``economics-oriented'' pricing for maximum revenue),
a great deal of scholarly work has explored the proper design of pricing schemes \cite{YU:2014}--\cite{Li:2014}.
However, a considerable mismatch between capacity and demand has been observed in reality \cite{Henry:2005}. 
One potential reason for such mismatch is that the {\it uncertainty} of network demand is largely ignored in the literature. 
For example, the network demand is often unknown to an operator in priori. 
Since the operator has to face or predict the demand uncertainty, it inevitably results in under-utilization of capacity.
In this work, we address this challenging issue, from a new economics-oriented pricing framework.

With the ever-increasing desire for more data volume,
network operators have widely adopted the usage-based pricing scheme in both wireline and wireless communication networks \cite{Dowell:2010, Cox:2010}. 
However, much of such pricing scheme is based on the monthly network demand, leading to high peak-to-average ratios of the network demand.
For example, the network demand in peak hours can be ten times more than that in off-peak hours \cite{Wong:2001}.
As a remedy to this inefficiency, 
time-dependent usage-based pricing (TDP) was introduced to reduce the peak-to-average ratios of 
the network demand, taking into the actual network demand \cite{Ha:2012}.
In \cite{Ha:2012}, the authors showed that 24\% of traffic can be redistributed over a day, through incentivizing users to shift their usage to off-peak hours with lower prices. 
Nevertheless, we argue that resorting to (operator-driven) TDP still poses the problem of matching current network capacity with the network demand, because TDP still has to face or predict the demand uncertainty, rather than embracing it.

As a complement to (operator-driven) TDP\footnote{For the rest of paper, we will use the term `forward pricing' to refer (operator-driven) TDP.}, 
in this paper, we propose {\it reverse pricing} to effectively handle unexpected demand fluctuations.
Exemplified by {\it Priceline.com} \cite{Spann:2010},
reverse pricing (also known as ``name-your-own-price'') is a market mechanism under which 
the network operator decides a resource recommendation rule and a hidden bid-acceptance threshold {\it a priori},
while each user submits a bid (unit price) that he is willing to pay for a given amount of resources. 
If the bid exceeds the unknown threshold price, it is accepted. The user then pays his bid and uses the specified resources.  
Intuitively, the superiority of reverse pricing is its natural ability to deal with the demand uncertainty over time. 
However, the challenge here is that such user participation in the pricing process may decrease the network operator's revenue, due
to the negative cannibalization effect. 
Then, a natural question is: How to design reverse pricing in conjunction with forward pricing subject to both demand uncertainty and capacity constraint, with the goal of boosting the network operator's expected revenue?

This paper endeavors to design such reverse pricing on top of forward pricing to answer the above question with stylized communication network models. 
To this end, we first model the interaction between a single network operator and the users as a four-stage Stackelberg game, 
when employing reverse pricing along with forward pricing.  
To investigate its economic viability,   
we propose the proportional residual resource recommendation rule in which 
a user is allocated the available remaining resources in proportion to his payment via forward pricing, 
and examine the hidden bid-acceptance threshold conditions under which it is more profitable to adopt reverse pricing.   
We analyze the case of forward pricing only, as a baseline to evaluate the effects of utilizing the reverse pricing scheme.
We show that our reverse pricing can achieve ``{\it triple-win}'' solutions: an increase in the average total revenue of the operator; higher average resource utilization efficiency; and an increment in the total average payoff of the users.

\begin{figure*}[t]
\centering
\includegraphics[angle=0,width=7.3in]{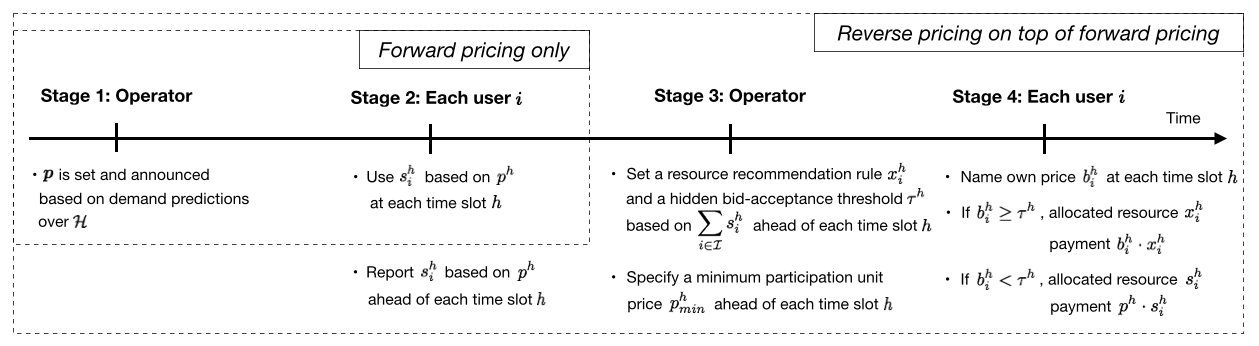}
\caption{Timing of the game.}
\end{figure*}

The rest of this paper is organized as follows:
Section II describes the system model. 
Section III introduces the reverse pricing mechanism and investigates its economic viability.
Section IV examines the forward pricing scheme only, 
as a baseline to evaluate the impact of the proposed reverse pricing scheme on the average total revenue of the 
operator's revenue, the users' payoffs, 
and the resource utilization efficiency.  
Section V gives numerical results to validate the proposed studies, followed by concluding remarks in Section VI.

\section{System Model}
To abstract the interaction between a single network operator and users, and to introduce our reverse pricing scheme smoothly, 
without loss of generality, we consider a specific system model hereafter. 
This, however, does not mean that our scheme is only applicable to such specific model. 
Now let us consider a communication network with a total amount of available resource $Q$ (e.g., bandwidth, data rate, data amount, etc). 
A single network operator allocates this resource to a set $\mathcal{I}\triangleq\{1,\cdots, I\}$ of users. 
We assume that the users have time-varying demand preferences for resources. 
These preferences are associated with their demand patterns over time.
In this regard, we consider a time-slotted system where the resource scheduling horizon is divided into a set $\mathcal{H}\triangleq\{1, \cdots, H\}$ of time slots. Note that this time slot can be based on time-varying demand patterns: peak network demand time slots, normal network demand time slots, and off-peak network demand time slots. 

Each user $i\in\mathcal{I}$ chooses a finite amount of resources based on his own time-varying demand preferences and the service prices over time. 
Specifically, user $i$ aims to maximize his own payoff, which is the difference between the utility of the allocated resource and the payment to the operator. 
Let $\boldsymbol{p}\triangleq\left[p^{h}\right]_{h\in\mathcal{H}}$ denote the (operator-driven) price vector per unit resource over the resource scheduling horizon $\mathcal{H}$. 
We assume $\boldsymbol{p}$ is known to the users before the resource scheduling horizon (i.e., ex-ante price), and remains constant at the end of the current scheduling horizon. 
Without loss of generality, the payoff function for user $i$ at time slot $h$ can be defined as 
\begin{equation}\small
u_{i}(\theta_{i}^{h}, s_{i}^{h}, p^{h})=\theta_{i}^{h}\ln(1+s_{i}^{h}) - p^{h}s_{i}^{h},\label{Eq:Payoff}
\end{equation}\normalsize
where $s_{i}^{h}$ is the allocated resource to user $i$, and $\theta_{i}^{h}$ is the time-varying willingness to pay of user $i$ per unit resource. 
As stated in \cite{Basar:2002, Jung:2013}, the logarithmic utility function in \eqref{Eq:Payoff} has been widely used to model the proportionally fair resource allocation in communication networks.

We consider two types of pricing schemes:   
\begin{enumerate}
\item \emph{Forward pricing only}: 
The operator publishes the ex-ante unit price vector $\boldsymbol{p}$ over the resource scheduling horizon $\mathcal{H}$. 
The purpose of the operator is to maximize its expected revenue based on demand predictions.
It implies that the operator has to encounter some level of {\it demand uncertainty}.
This residual uncertainty stemming from users' actual resource consumption may waste some available resources.

\item \emph{Reverse pricing on top of forward pricing}: 
As a complement to forward pricing, 
the operator partially relinquishes its pricing control to the users. 
However, we assume that each user $i$ needs to report the amount of resources $s_{i}^{h}$ based on $p^{h}$, {\it ahead of} time slot $h$. 
From this contract, the operator calculates the residual resources $Q-\sum_{i\in\mathcal{I}}s_{i}^{h}$, 
and suggests to allocate more resources to user $i$ in proportional to $s_{i}^{h}$, 
i.e., $x_{i}^{h} = s_{i}^{h} + \frac{s_{i}^{h}}{\sum_{j\in\mathcal{I}}s_{j}^{h}}(Q - \sum_{j\in\mathcal{I}}s_{j}^{h})$.
Under this proportional residual resource recommendation rule, 
the operator publicly specifies the minimum participation unit price $p_{min}^{h}$ below which $x_{i}^{h}$ is not sold before time slot $h$, and 
generates the hidden bid-acceptance threshold $\tau^{h}$ (price) with a known distribution on the interval $[p_{min}^{h}, p^{h})$,
 Specifically, $\tau^{h}$ is unknown to the users, but the distribution is common knowledge among the users. 
Then, user $i$ can name his own unit price $b_{i}^{h}$ at will for $x_{i}^{h}$ and receives $x_{i}^{h}$ if $b_{i}^{h}$ exceeds $\tau^{h}$ at each time slot $h$ (If not, user $i$ receives $s_{i}^{h}$ with the unit price $p^{h}$). 
\end{enumerate}

Fig. 1 shows the timing of the game that characterizes the interaction between the operator and users. 
In Stage I, the operator announces the unit price vector to the users over the resource scheduling horizon, 
estimating the potential total demand from users. 
Due to the limited resource available, 
the operator should design the forward pricing scheme to guarantee that the aggregate user demand is no larger than what it can provide. 
In Stage II, each user acts as a price taker and chooses the amount of resources over time according to the published unit price vector. 
When the operator employs forward pricing only, the game ends. 
When the operator adopts reverse pricing on top of forward pricing, on the other hand, 
each user is required to report his resource usage via forward pricing, before the beginning of each time slot. 
In Stage III, 
the operator calculates the total demand from the users via forward pricing, 
sets a resource recommendation rule, and generates a hidden bid-acceptance threshold with a minimum participation unit price via reverse pricing. 
In Stage IV, each user names his own unit price for the given amount of resources, 
and the resource is allocated to him immediately and accordingly depending on whether or not his submitted bid exceeds the operator's hidden bid-acceptance threshold. 
We solve these two different games by using backward induction from Stage IV to Stage I, 
and discuss one by one in the following sections.

\section{Reverse Pricing On Top Of Forward Pricing}

We first consider that the operator adopts the reverse pricing scheme on top of the forward pricing scheme. 
Given the (operator-driven) unit price vector and the corresponding total demand of the users over the resource scheduling horizon, 
we investigate the users' participation decisions and corresponding bidding strategies, 
as well as the operator's resource recommendation rule and hidden bid-acceptance threshold with a known distribution. 
We start with the users's cases, with the hope to shed light on the feasibility of utilizing the reverse pricing scheme.

\subsection{Users' Participation Decisions and Subsequent Bidding Strategies in Stage IV}

Without loss of generality, 
we assume that $\boldsymbol{x}_{i}\triangleq\left[x_{i}^{h}\right]_{h\in\mathcal{H},i\in\mathcal{I}}$ denote the user $i$'s resource vector declared by the operator in Stage III. 
For analytical tractability, we assume that the operator generates the hidden bid-acceptance threshold $\tau^{h}$ that is uniformly distributed 
on the interval $[p_{min}^{h}, p^{h}]$.
We remark that $0\leq p_{min}^{h} < p^{h}$ is the {\it minimum participation unit price}, where each user should submit his bid at least $p_{min}^{h}$ at time slot $h$ via reverse pricing.

Given $x_{i}^{h}$ and $p_{min}^{h}$ at time slot $h$, user $i$ decides whether to participate in the pricing process, i.e.,  
\begin{equation}\small
\thickmuskip=2mu\medmuskip=1mu
\theta_{i}^{h}\ln(1+x_{i}^{h}) - p_{min}^{h}x_{i}^{h}\geq \theta_{i}^{h}\ln(1+s_{i}^{h}) - p^{h}s_{i}^{h}, \forall i\in\mathcal{I}, \forall h\in\mathcal{H}, \label{Eq:p_{min}}
\end{equation}\normalsize
where $s_{i}^{h}$ is the user $i$'s reported resource usage in Stage II, based on the announced price per unit resource $p^{h}$ in Stage I.

For inducing user $i$ to take part in the pricing process, 
$p_{min}^{h}$ must satisfy the constraint \eqref{Eq:p_{min}}, i.e., 
\begin{equation}\small
p_{min}^{h}\leq \frac{\theta_{i}^{h}\ln\left((1+x_{i}^{h})/(1+s_{i}^{h})\right)+p^{h}s_{i}^{h}}{x_{i}^{h}}.
\end{equation}\normalsize
Otherwise, user $i$ has no incentive to name his own unit price $b_{i}^{h}$ (i.e., $b_{i}^{h}=0$).
Based on it, we define $\mathcal{I}^{+}(p_{min}^{h})=\left\{i\in\mathcal{I}: p_{min}^{h}\leq \frac{\theta_{i}^{h}\ln\left((1+x_{i}^{h})/(1+s_{i}^{h})\right)+p^{h}s_{i}^{h}}{x_{i}^{h}}\right\}$, i.e., the set of users that would name strictly positive prices at time slot $h$ in Stage IV.

Since each user $i\in\mathcal{I}^{+}(p_{min}^{h})$ is uncertain to the hidden bid-acceptance threshold $\tau^{h}$,  
each user $i$ names his own price $b_{i}^{h}$ to maximize the following expected payoff:

\small\begin{eqnarray}
\medmuskip=-3mu \thickmuskip=-1mu
{\bf{P1}}:\max_{p_{min}^{h}\leq b_{i}^{h}\leq p^{h}}\;\;u_{i}(\theta_{i}^{h}, x_{i}^{h}, b_{i}^{h})\mathrm{P}(b_{i}^{h}\geq \tau^{h})+
\quad\quad\;\quad\quad\quad\;\;\;\;\nonumber 
\\ 
 u_{i}(\theta_{i}^{h}, s_{i}^{h}, p^{h})\mathrm{P}(b_{i}^{h} < \tau^{h}), \forall i\in\mathcal{I}^{+}(p_{min}^{h}), \forall h\in\mathcal{H},\label{Eq:Bid}
\end{eqnarray}\normalsize

\emph{Proposition 1:} For each user $i\in\mathcal{I}$,
the optimal solution for Problem {\bf{P1}} at each time slot $h\in\mathcal{H}$ is given by 
\begin{equation}\small
\medmuskip=-1mu \thickmuskip=1mu
b_{i}^{h^{\ast}}=
\begin{cases}
\frac{1}{2x_{i}^{h}}\left[\theta_{i}^{h}\ln\left(\frac{1+x_{i}^{h}}{1+s_{i}^{h}}\right)+s_{i}^{h}p^{h}+x_{i}^{h}p_{min}^{h}\right], &\forall i\in\mathcal{I}^{+}(p_{min}^{h}), \\
0, &\forall i\in\mathcal{I}\backslash\mathcal{I}^{+}(p_{min}^{h}). \label{Eq:OptBid}
\end{cases}
\end{equation}\normalsize
\begin{proof}
To verify that Problem {\bf{P1}} is a convex optimization with respect to $b_{i}^{h}$, let us rewrite this problem as follows:
\vskip -10pt
\small\begin{equation*}
\max_{p_{min}^{h}\leq b_{i}^{h}\leq p^{h}} \left(\theta_{i}^{h}\ln(1+x_{i}^{h})-b_{i}^{h}x_{i}^{h}\right)\frac{b_{i}^{h}-p_{min}^{h}}{p^{h}-p_{min}^{h}} +
\quad\quad\quad\quad\quad 
\end{equation*}
\normalsize
\begin{equation}\small
\medmuskip=1mu \thickmuskip=1mu
\left(\theta_{i}^{h}\ln(1+s_{i}^{h})-p_{i}^{h}s_{i}^{h}\right)\frac{p^{h}-b_{i}^{h}}{p^{h}-p_{min}^{h}}, \forall i\in\mathcal{I}^{+}(p_{min}^{h}), \forall h\in\mathcal{H}.
\end{equation}\normalsize

Apparently, the constraint is a convex set. Thus, it suffices to examine that the second order derivative is less than 0.  
Let $f_{i}(b_{i}^{h})$ denote the user $i$'s objective function in \eqref{Eq:Bid}. Then we have 
\medmuskip=3mu \thickmuskip=3mu \thinmuskip=3mu
\begin{equation}\small
\frac{\partial^{2}f_{i}}{\partial b_{i}^{h^{2}}} = \frac{2x_{i}^{h}}{p^{h}-p_{min}^{h}} < 0,
\end{equation}\normalsize
which proves that Problem {\bf{P1}} is a convex optimization with respect to $b_{i}^{h}$. Exploiting the first order necessary condition yields
\begin{equation}\small
b_{i}^{h^{\ast}}=\frac{1}{2x_{i}^{h}}\left[\theta_{i}^{h}\ln\left(\frac{1+x_{i}^{h}}{1+s_{i}^{h}}\right)+s_{i}^{h}p^{h}+x_{i}^{h}p_{min}^{h}\right], 
\end{equation}\normalsize
which completes the proof. 
\end{proof}

Proposition 1 characterizes the optimal bidding strategies of the users in Stage IV. 
Since the users maximize their expected payoff against the probability of being accepted via reverse pricing, 
they tend to name their own unit prices below the unit price that the operator sets 
(i.e., $b_{i}^{h^{\ast}} =(p_{min}^{h}+p^{h})/2 < p^{h}$ when $x_{i}^{h}=s_{i}^{h}$). 
It hints on how to design the reverse pricing mechanism to extract more revenue from the users.

\subsection{Operator's Resource Recommendation Rule and Hidden Bid-Acceptance Threshold in Stage III}

We now consider how to utilize the reverse pricing scheme along with the forward pricing scheme. 
To this end, we are interested in examining the following two issues:
\begin{enumerate}
\item 
How to design a resource recommendation rule based on the total demand from the users in Stage II, subject to the total limited resource? 
\item 
How to commit to a minimum markup strategy for generating the hidden bid-acceptance threshold 
with the goal of boosting the operator's expected revenue?
\end{enumerate}

To address these issues one by one, we first investigate the resource recommendation rule in the reverse pricing scheme. 
Let $\boldsymbol{Q_{r}}\triangleq\left[\left(Q-\sum_{i\in\mathcal{I}}s_{i}^{h}\right)^{+}\right]_{h\in\mathcal{H}}$ denote the residual resource available
over the resource scheduling horizon $\mathcal{H}$, where $\left(\cdot\right)^{+}$ denotes $\max\{\cdot,0\}$.
In the special case where no resource is left to the operator at time slot $h$ (i.e., $Q_{r}^{h}=0$), 
the operator allocates the same amount of resources submitted by the users in Stage II (i.e., $x_{i}^{h}=s_{i}^{h}$).

When $Q_{r}^{h}>0$, on the other hand, the operator decides the resource recommendation rule to satisfy the following conditions:
\begin{equation}\small
Q\geq \sum_{i\in\mathcal{I}}x_{i}^{h}\geq \sum_{i\in\mathcal{I}}s_{i}^{h}, \quad \forall h\in\mathcal{H}\label{Eq:constraint}
\end{equation}\normalsize
where the first inequality in \eqref{Eq:constraint} corresponds to the maximum available resource constraint that the operator can provide via reverse pricing, and the second inequality in \eqref{Eq:constraint} comes from the operator's revenue maximization standpoint through the disposal of residual available resources, as noted in the previous subsection.   
In this case, intuitively, the operator seeks to clear the market, i.e., $Q=\sum_{i\in\mathcal{I}}x_{i}^{h}$ for revenue maximization.

Capturing the idea that more resources reported from the users in Stage II implicitly mean higher willingness to pay of them, 
in this work, we propose the proportional residual resource recommendation rule. 
To be more specific, the operator simply suggests to allocate the remaining amount of resources to each user $i$ in proportion to his reported amount of resources $s_{i}^{h}$ in Stage II (i.e., $x_{i}^{h^{\ast}}=s_{i}^{h}+\frac{s_{i}^{h}}{\sum_{j\in\mathcal{I}}s_{j}^{h}}Q_{r}^{h}$).

Next we explain how the operator determines a hidden bid-acceptance threshold $\tau^{h}$ 
given the proportional residual resource recommendation rule. 
To this end, 
the operator should examine the conditions under which it is more profitable to employ the reverse pricing scheme.  
The answer to these conditions is intertwined with the minimum participation unit price $p_{min}^{h}$.

Given that user $i$ reports his resource consumption $s_{i}^{h}$ via forward pricing in Stage II, 
the operator tries to set $p_{min}^{h}$, i.e., 
\begin{equation}\small
p_{min}^{h}x_{i}^{h^{\ast}}\geq p^{h}s_{i}^{h}, \quad \forall i\in\mathcal{I}, \forall h\in\mathcal{H}, \label{Eq:rev}
\end{equation}\normalsize
where the constraint \eqref{Eq:rev} indicates that the operator avoids the revenue loss via reverse pricing, 
by setting $p_{min}^{h}$ to earn at least the revenue via forward pricing.
We then have the following results as shown in Lemma 1.

\emph{Lemma 1:} Assume that an operator adopts a proportional residual resource recommendation rule based on the reported aggregate user demand in Stage II. 
The minimum participation unit price that the operator sets should satisfy the following condition
\begin{equation}\small
p^{h}> p_{min}^{h}\geq \frac{p^{h}\sum_{i\in\mathcal{I}}s_{i}^{h}}{Q}, \quad\forall h\in\mathcal{H}. \label{Eq:minimum bid}
\end{equation}\normalsize
\begin{proof}
The proportional residual resource recommendation rule leads to $Q=\sum_{i\in\mathcal{I}}x_{i}^{h^{\ast}}$. 
Summing over all $i$ under the constraint \eqref{Eq:rev} completes the proof. 
\end{proof}

Lemma 1 shows the {\it viability} of utilizing the reverse pricing scheme on top of the forward pricing scheme.
By exploiting the revealed total user demand in Stage II, the operator can boost its revenue. 

For revenue maximization, however, 
the operator should further investigate its expected revenue under different values of $p_{min}^{h}$. 
When $p_{min}^{h}$ is small, in general, the users tend to bid with lower prices (see the last term in Eq. \eqref{Eq:OptBid}).
This implies that more potentially profitable trades are expected to occur, 
at the cost of the higher decrease in its revenue per user via reverse pricing.
When $p_{min}^{h}$ is large, on the other hand, the reverse is generally true.  
Thus, the operator should set $p_{min}^{h}$ as a {\it strategic} variable, taking this trade-off into account. 
A numerical example of this will be discussed in Sec. V.

\section{Benchmark Scenario: Forward Pricing Only}
In the previous section, 
we have investigated the viability of introducing the reverse pricing scheme on top of the forward pricing scheme. 
In this section, we consider the case of the forward pricing scheme only. This study serves as a baseline to evaluate the impact and significance of the reverse pricing scheme on the average operator's revenue, the average total user payoff, and the average resource utilization efficiency. 
By backward induction again, we start from Stage II where each user acts as a price taker and adjusts his amount of resource over time.

\subsection{Users' Desired Resources in Stage II}

As described in Sec. II, 
if the operator announces a unit price $p^{h}$ at each time slot $h\in\mathcal{H}$ in Stage I, 
each user $i\in\mathcal{I}$ solves the following payoff maximization problem:
\begin{equation}\small
{\bf{P2}}:\max_{s_{i}^{h}}u_{i}(\theta_{i}^{h}, s_{i}^{h}, p^{h})=\theta_{i}^{h}\ln(1+s_{i}^{h}) - p^{h}s_{i}^{h},\label{Eq:MaxPayoff}
\end{equation}\normalsize
which leads to 
\begin{equation}\small
s_{i}^{h^{\ast}}=\left(\frac{\theta_{i}^{h}}{p^{h}}-1\right)^{+}.
\end{equation}\normalsize

In the following analysis, throughout the paper, we only focus on a special case of Problem ${\bf{P2}}$ by assuming that 
all the users are always admitted (i.e., $s_{i}^{h^{\ast}}>0, \forall i\in\mathcal{I}, \forall h\in\mathcal{H}$)\footnote{This assumption does not change the main insights obtained in this paper.}.

\subsection{Operator's Forward Pricing in Stage I}
In Stage I, the operator tries to maximize its expected
revenue over the resource scheduling horizon based on demand
predictions.
For the operator experiencing a variety of time-varying demand patterns,
the assumption of the complete information of the network demand a priori is not valid in general. 
To limit the dimension of this issue, 
we assume that the operator knows the payoff function of each user $i$ in \eqref{Eq:Bid}, 
but $\theta_{i}^{h}$ is a random variable with a finite mean with a bounded magnitude at each time slot $h$.
For the ease of exposition, the time-varying willingness to pay for user $i$ can be expressed as $\theta_{i}^{h}= \tilde{\theta}_{i}^{h} + \delta_{i}^{h}$, where $\tilde{\theta}_{i}^{h}$ is a finite mean and $\delta_{i}^{h}$ is a random variable that indicates a time variance in
user $i$'s willingness to pay. We further assume that $\delta_{i}^{h}$ is a zero-mean random variable with a bounded magnitude known to the operator. Mathematically, $\mathrm{E}\left(\delta_{i}^{h}\right)=0$ and $|\delta_{i}^{h}|\leq \epsilon_{i}^{h}$, with $\epsilon_{i}^{h}>0$ denoting the maximum magnitude of user $i$'s demand uncertainty, $\forall i\in\mathcal{I}, \forall h\in\mathcal{H}$.
Without loss of generality, 
we assume that $\tilde{\theta_{1}^{h}}-\epsilon_{1}^{h}\geq \tilde{\theta}_{2}^{h}-\epsilon_{2}^{h}\geq\cdots\geq\tilde{\theta}_{I}^{h}-\epsilon_{I}^{h}$ at each time slot $h$.

Under the above assumption, 
the operator seeks to announce the optimal price vector $\boldsymbol{p}$ to maximize its expected revenue over the resource scheduling horizon $\mathcal{H}$,
subject to the limited total resource $Q$. 
This is obtained by solving the following optimization problem:
\begin{equation}\small
{\bf{P3}}:\max_{\boldsymbol{p\geq 0}}\sum_{h\in\mathcal{H}}\sum_{i\in\mathcal{I}}\mathrm{E}_{\theta_{i}^{h}}\left(\theta_{i}^{h}-p^{h}\right),\quad\quad\quad\quad \label{Eq:ForwardPrice}
\end{equation}\normalsize
\vskip -5pt
\begin{equation}\small
\quad\quad\quad\text{s.t.}\sum_{i\in\mathcal{I}}\left(\frac{\theta_{i}^{h}}{p^{h}}-1\right)\leq Q, \quad\forall h\in\mathcal{H}. \label{Eq:ForwardConstraint}
\end{equation}\normalsize
Note that we do not take the expectation of the resource constraint in \eqref{Eq:ForwardConstraint} with respect to $\theta_{i}^{h}$,
to always guarantee that the aggregate user demand in Stage II is no larger than what the operator can provide.

For satisfying the resource constraint in \eqref{Eq:ForwardConstraint}, 
the operator should set the ex-ante price $p^{h}$, i.e., 
\begin{equation}\small
\sum_{i\in\mathcal{I}}\left(\frac{\theta_{i}^{h}}{p^{h}}-1\right)\mathop \leq \limits^{(a)} \sum_{i\in\mathcal{I}}\left(\frac{\tilde{\theta}_{i}^{h}+\epsilon_{i}^{h}}{p^{h}}-1\right)\leq Q, \;\forall h\in\mathcal{H},
\end{equation}\normalsize
where $(a)$ follows from the maximum value of $\theta_{i}^{h}$. 
Since the objective function in \eqref{Eq:ForwardPrice} is a decreasing function of $p^{h}$ at each time slot $h$,
we can reformulate Problem ${\bf{P3}}$ as follows: 
\begin{equation}\small
{\bf{P4}}:\min_{\boldsymbol{p\geq 0}}\sum_{h\in\mathcal{H}}\sum_{i\in\mathcal{I}}p^{h},\quad\quad\quad\quad\quad\quad\quad\quad\quad\quad\quad \label{Eq:MinOb}
\end{equation}\normalsize
\vskip -5pt
\begin{equation}\small
\quad\quad\text{s.t.}\sum_{i\in\mathcal{I}}\left(\frac{\tilde{\theta}_{i}^{h}+\epsilon_{i}^{h}}{p^{h}}-1\right)\leq Q, \quad\forall h\in\mathcal{H}. \label{Eq:MinCon}
\end{equation}\normalsize

\emph{Proposition 2:}
Given $Q>\frac{\sum_{i\in\mathcal{I}}\tilde{\theta}_{i}^{h}+\epsilon_{i}^{h}}{\tilde{\theta}_{I}^{h}-\epsilon_{I}^{h}}-I$, 
the optimal solution to Problem ${\bf{P4}}$ is given by
\begin{equation}\small
p^{h^{\ast}}=\frac{\sum_{i\in\mathcal{I}}\tilde{\theta}_{i}^{h}+\epsilon_{i}^{h}}{Q+I}, \quad\forall h\in\mathcal{H}. \label{Eq:OptForwardPrice}
\end{equation}\normalsize 
\begin{proof}
The objective function in \eqref{Eq:MinOb} is an increasing function of $p^{h}$ at time slot $h$. On the other hand, the left hand side of the resource constraint in \eqref{Eq:MinCon} is decreasing in $p^{h}$. Thus the optimal solution can be obtained when the equality in \eqref{Eq:MinCon} holds. 
Since all the users are always admitted, the optimal solution \eqref{Eq:OptForwardPrice} in should be less than $\tilde{\theta}_{I}^{h}-\epsilon_{I}^{h}$, which completes the proof.  
\end{proof}

Proposition 2 reveals the practical limitation, from the operator's revenue maximization perspective.
As the level of {\it demand uncertainty} $\epsilon_{i}^{h}$ is higher, the operator has to charge a higher price $p_{i}^{h^{\ast}}$ to the users
to satisfy the resource constraint in \eqref{Eq:MinCon}. It may lead to large demand fluctuations at the operator in accordance with 
users' actual resource consumption.

\begin{figure}[t]
\centering
 \subfigure[Average network demand]{
   \includegraphics[angle=0, width=3.5in] {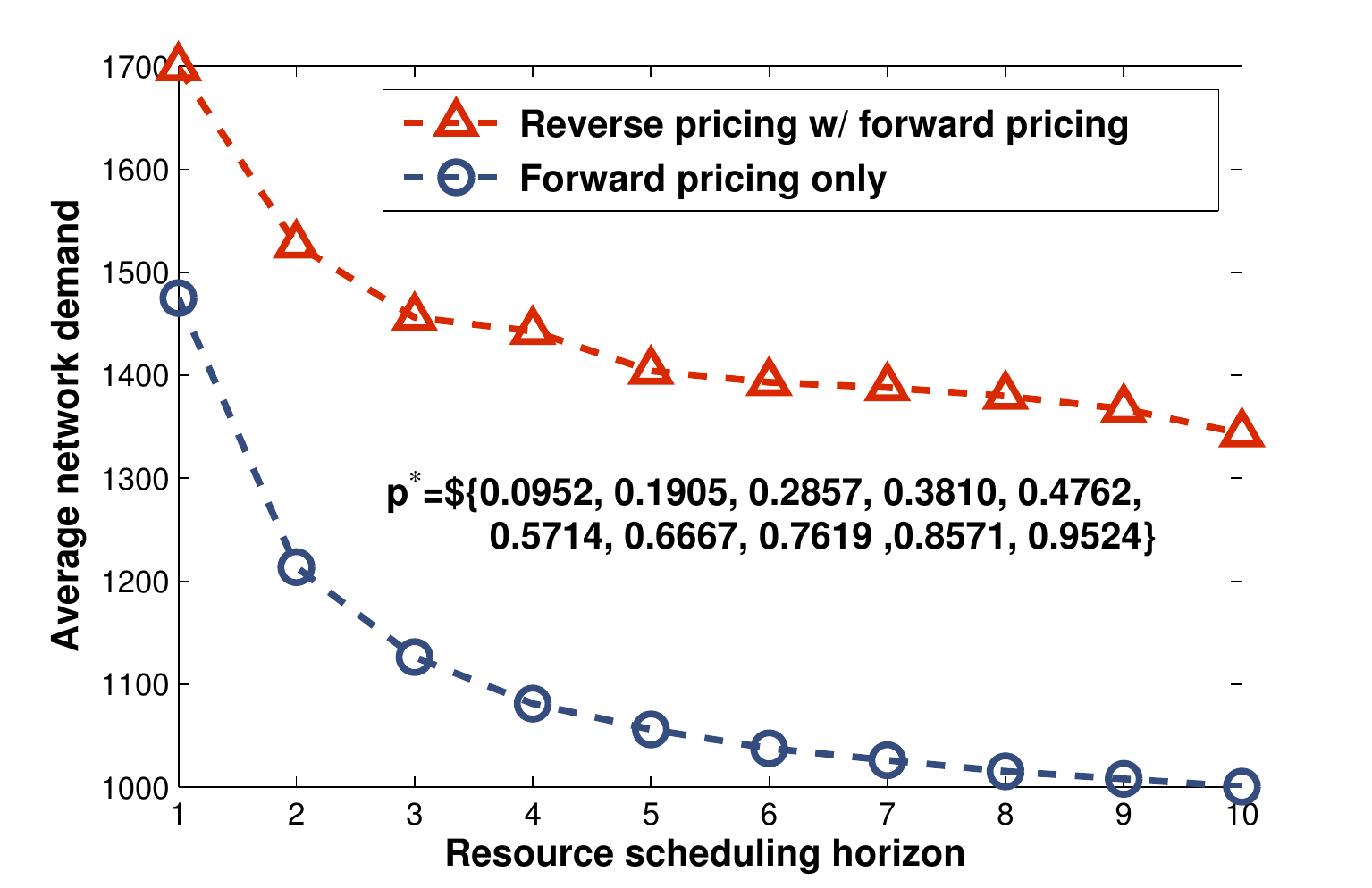}
   \label{fig:subfig1}
 }
\subfigure[Average total user payoff]{
   \includegraphics[angle=0,width=3.5in] {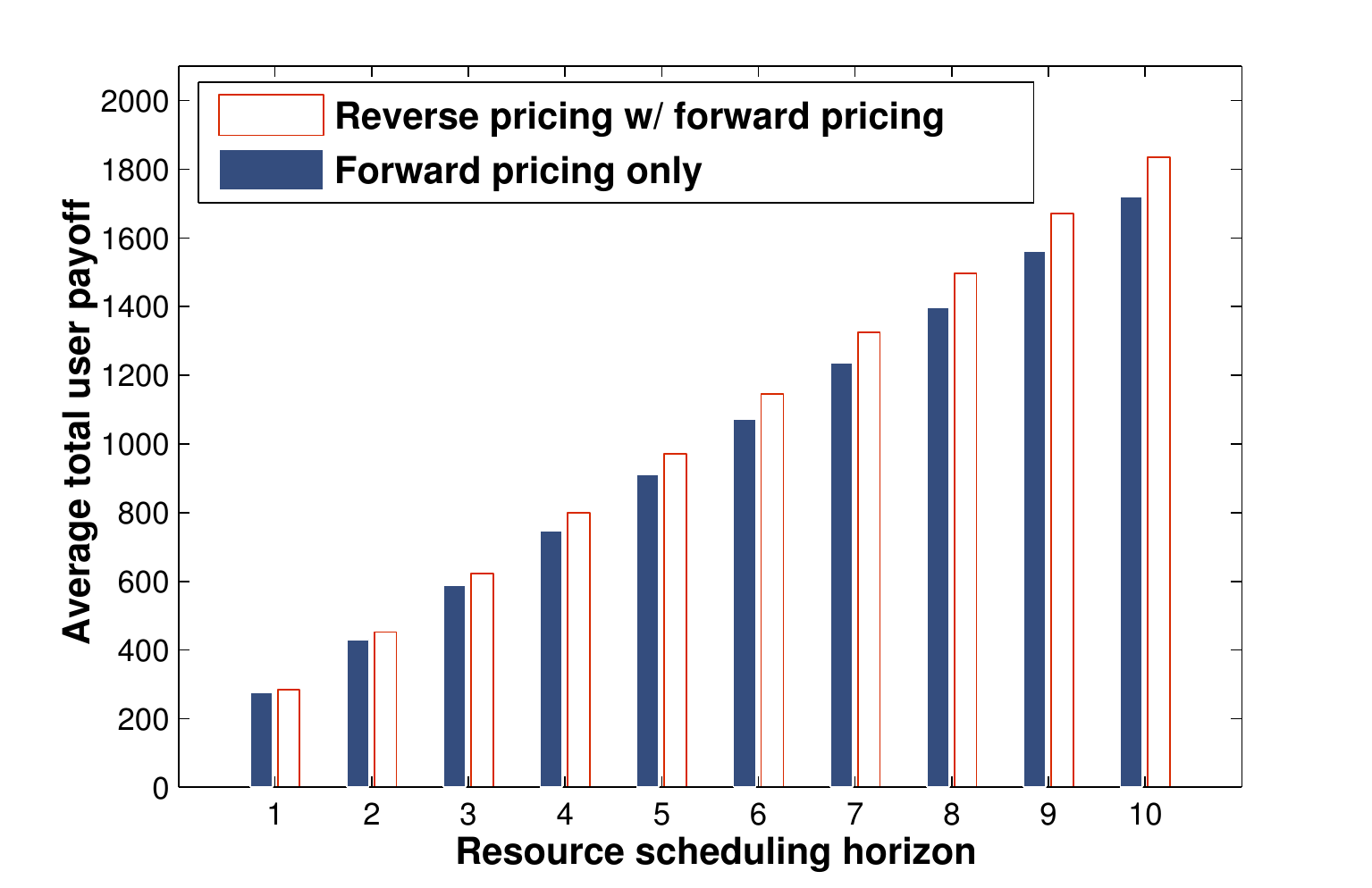}
   \label{fig:subfig2}
 }
\subfigure[Average operator's revenue]{
   \includegraphics[angle=0,width=3.5in] {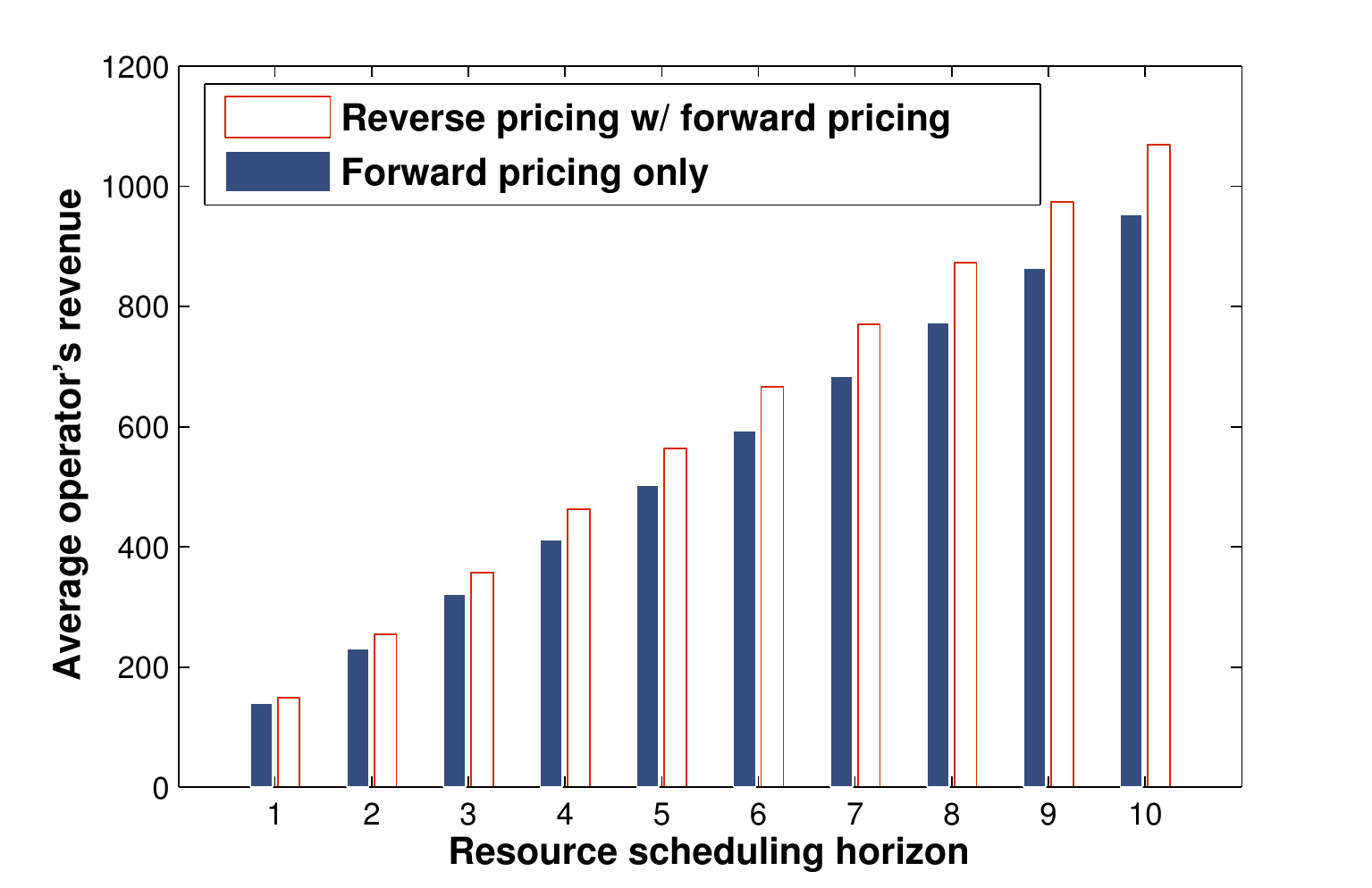}
   \label{fig:subfig3}
 } 
\caption{Reverse pricing on top of forward pricing vs. forward pricing only.}
\end{figure}

\section{Numerical Results}
We provide numerical examples to study several key properties of reverse pricing on top of forward pricing.
Consider a 100-user in the network where the total amount of available resource is chosen as $Q=1000$.
The resource scheduling horizon is set as $\mathcal{H}=\{1, 2, \cdots, 10\}$.
The time-varying willingness to pay of user $i$ follows the uniform distribution on $[1, 2h]$ at each time slot $h$.
The minimum participation unit price is fixed as $p_{min}^{h}=\frac{p^{h}\sum_{i\in\mathcal{I}}s_{i}^{h}}{Q}$ unless specified otherwise.
Each graph represents an average over 1000 independent realizations at each time slot $h$. 

Fig. 2(a) shows how the average network demand changes over the resource scheduling horizon. 
With forward pricing only, 
there is a still large difference between the maximum available resource and the average network demand.   
In this simulation, as each time slot passes over the resource scheduling horizon, such difference becomes more distinct due to the higher level of demand uncertainty. 
With reverse pricing on top of forward pricing, on the other hand, it is more effectively handled.
From this, 
it is observed from Fig. 2(b) that our proposed pricing scheme always outperforms the forward pricing scheme only, 
in terms of the average total user payoff. Even if each user reports his actual resource usage via forward pricing, 
the incentive to name his own price brings benefits to him.
Fig. 2(c) shows that the operator can extract more revenue with reverse pricing in conjunction with forward pricing. 
Rather than facing the demand uncertainty, 
this result implies that the operator should re-examine the idea of involving the users in pricing decisions.

We then examine the effects of the minimum participation unit price level $p_{min}^{5}/p^{5^{\ast}}$ at a particular time slot (i.e., time slot 5).
Figs. 3(a) and 3(b) show that the average network demand and the average total user payoff are both non-increasing over $p_{min}^{5}/p^{5^{\ast}}$. This is due to the fact that the users are induced to submit higher bids as $p_{min}^{5}/p^{5^{\ast}}$ increases. 
Fig. 3(c) shows the average operator's revenue as a function of $p_{min}^{5}/p^{5^{\ast}}$.
When $p_{min}^{5}/p^{5^{\ast}}$ is low (i.e., $p_{min}^{5}/p^{5^{\ast}} \leq 0.2$),
the operator can obtain more average revenue via forward pricing only (i.e., negative cannibalization effect). 
When $p_{min}^{5}/p^{5^{\ast}}$ is intermediate (i.e., $0.2< p_{min}^{5}/p^{5^{\ast}} < 0.9$), 
the operator can achieve more average revenue via reverse pricing on top of forward pricing (i.e., positive cannibalization effect). 
With the goal of maximizing its average revenue, $p_{min}^{5}/p^{5^{\ast}}$ is chosen as 0.7, achieving a large revenue gain (e.g., 14\%) compared to forward pricing only. 
When $0.9\leq p_{min}^{5}/p^{5^{\ast}} \leq 1$, no users name their own prices, degenerating the forward pricing only case.

\section{Concluding Remarks}

\begin{figure}[t]
\centering
 \subfigure[Average network demand]{
   \includegraphics[angle=0, width=3.5in] {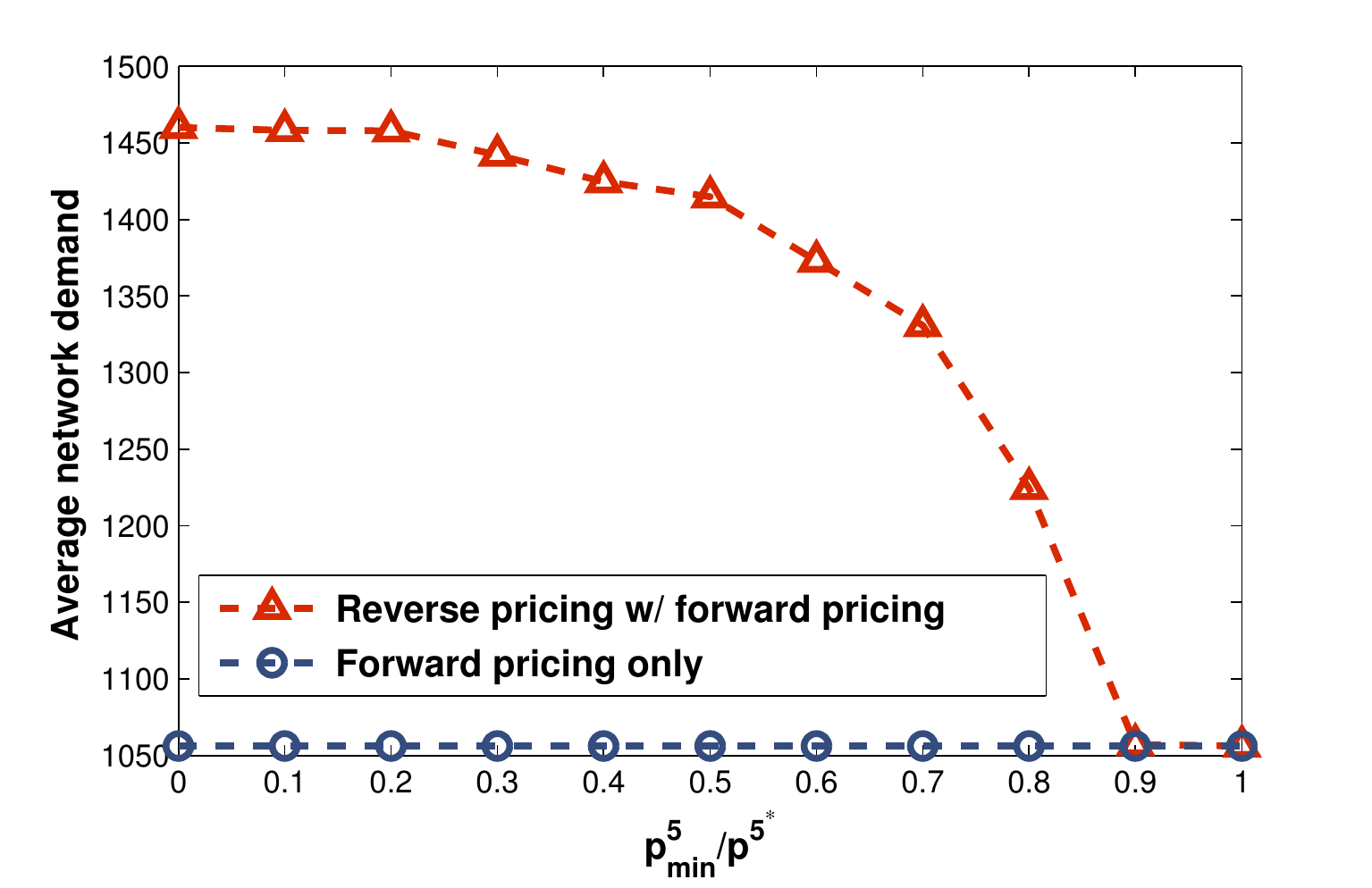}
   \label{fig:subfig1}
 }
\subfigure[Average total user payoff]{
   \includegraphics[angle=0,width=3.5in] {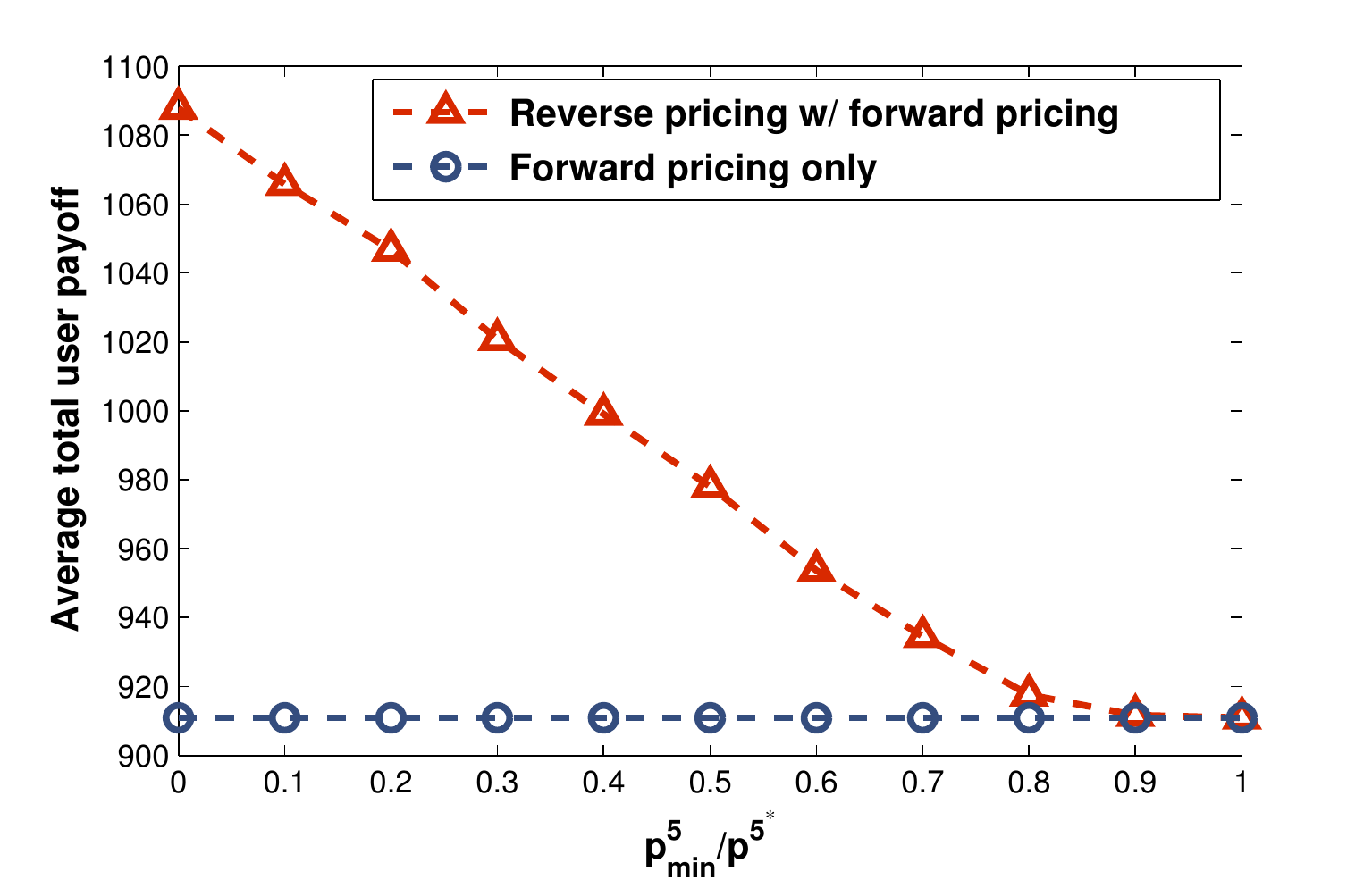}
   \label{fig:subfig2}
 }
\subfigure[Average operator's revenue]{
   \includegraphics[angle=0,width=3.5in] {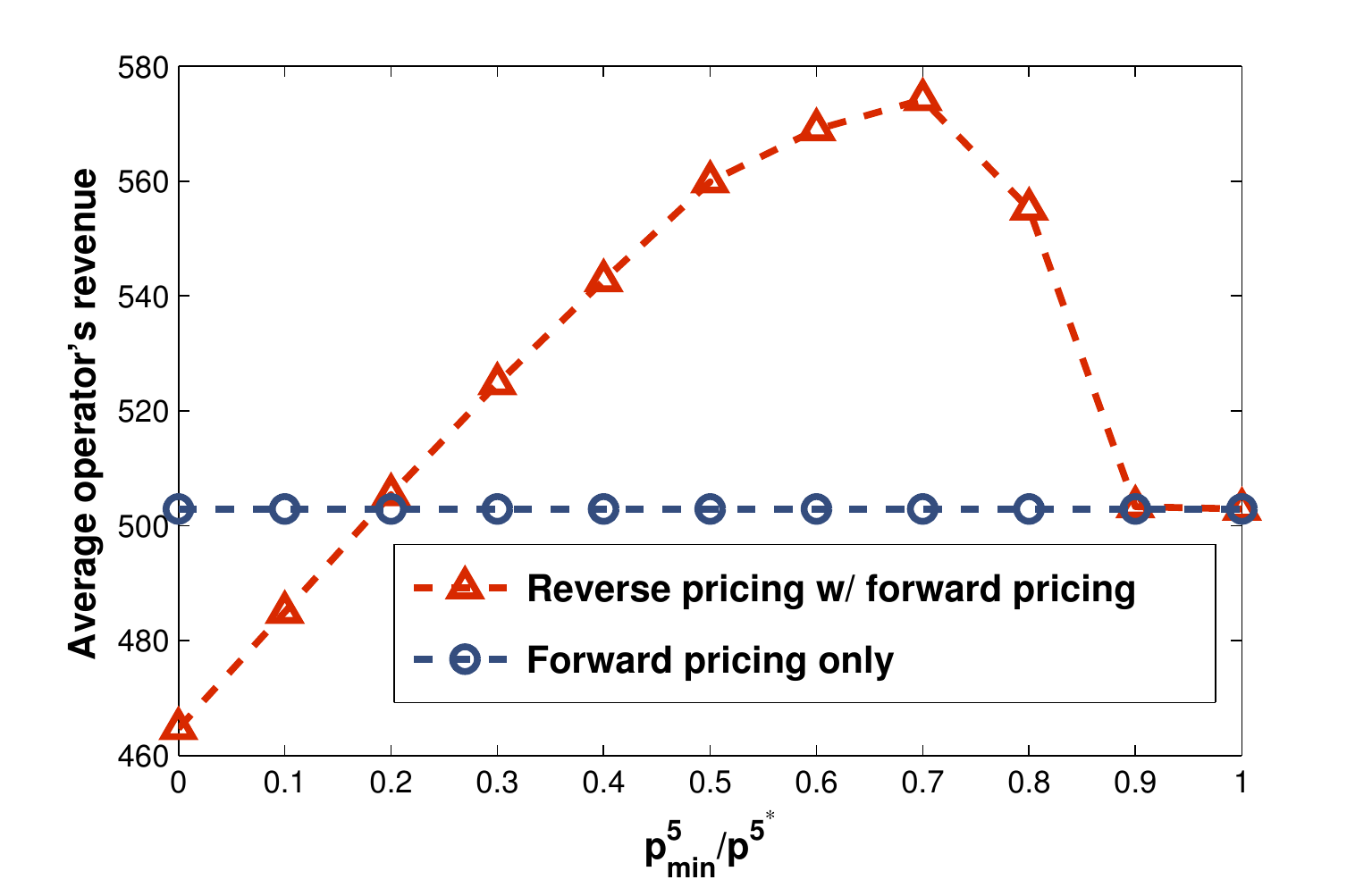}
   \label{fig:subfig3}
 } 
\caption{Impact of the minimum participation unit price level ($p_{min}^{5}/p^{h^{\ast}}$) on the average network demand, the average total user payoff, and the average operator's revenue at time slot 5.}
\end{figure}

This paper studies resource allocation mechanisms with reverse pricing for communication networks.  
As a complement to forward pricing, we design the reverse pricing mechanism to effectively handle unexpected demand fluctuations.
We show that the proposed pricing scheme can obtain 
``{\it triple-win}'' solutions: an increase in the average total revenue of the operator; higher average resource utilization efficiency; and an increment in the total average payoff of the users. In communication networks, pricing has been used for congestion-control, where the problem is a utility maximization while satisfying capacity constraints. This problem has been a classical issue since the work of Kelly \cite{Kelly98}. Therefore, we believe our reverse pricing will change the structure of 
the traditional congestion-control that are based on price-signaling depending on the congestion levels (i.e., Lagrangian multipliers). 
This will be an interesting issue where our current research is heading.

\section*{Acknowledgement}
This research was supported by the MSIP(Ministry of Science, ICT and Future Planning), Korea, under the CPRC(Communication Policy Research Center) support program (IITP-2015-H8201-15-1003) supervised by the IITP(Institute for Information \& communications Technology Promotion).

\end{document}